\documentclass[twocolumn,prb,floats,superscriptaddress]{revtex4}
\usepackage{graphicx,epsfig}
\usepackage{times}
\usepackage{graphics,dcolumn,bm,float}
\usepackage{amssymb,amsmath,rotate,color}
\usepackage{amsmath}

\begin{document}
\unitlength 1 cm
\newcommand{\be}{\begin{equation}}
\newcommand{\ee}{\end{equation}}
\newcommand{\bearr}{\begin{eqnarray}}
\newcommand{\eearr}{\end{eqnarray}}
\newcommand{\nn}{\nonumber}
\newcommand{\vk}{\vec k}
\newcommand{\vp}{\vec p}
\newcommand{\vq}{\vec q}
\newcommand{\vkp}{\vec {k'}}
\newcommand{\vpp}{\vec {p'}}
\newcommand{\vqp}{\vec {q'}}
\newcommand{\bk}{{\bf k}}
\newcommand{\bp}{{\bf p}}
\newcommand{\bq}{{\bf q}}
\newcommand{\br}{{\bf r}}
\newcommand{\bR}{{\bf R}}
\newcommand{\up}{\uparrow}
\newcommand{\down}{\downarrow}
\newcommand{\fns}{\footnotesize}
\newcommand{\ns}{\normalsize}
\newcommand{\cdag}{c^{\dagger}}
\newcommand{\degree}{\ensuremath{^\circ}}

\title{Crystalline anisotropy induces a second antiferromagnetic phase in the absence of SDW in the heavily hydrogen-doped  LaFeAsO$_{1-x}$H$_x$  $(x\sim0.5 )$ }
\author{Mehdi Hesani}\email{mehdi.hesani@gmail.com}
\author{Ahmad Yazdani}
\author{Kourosh Rahimi}
\affiliation{Condensed Matter Group, Department of Basic Sciences, Tarbiat Modares University, Jalal-Ale-Ahmad Avenue, Tehran, Iran }

\date{\today}

\begin{abstract}
Electronic and magnetic properties of the heavily H-doped LaFeAsO$_{1-x}$H$_x$  $(x\sim0.5 )$ were studied in the framework of the density functional theory combined with the dynamical mean field theory (DFT+DMFT). We found a stripe-like-ordered structure of hydrogen and oxygen atoms, as a ground state, with the same configuration as the antiferromagnetic (AF) order. The new configuration could explain the existing experimental results related to the heavily H-doped LaFeAsO$_{1-x}$H$_x$, such as an in-plane electronic anisotropy and a non-uniform magnetic behavior. A significant anisotropy was observed between Fe- 3d$_{xz}$ (xz) and Fe-3d$_{yz}$ (yz) orbitals in the ground state in the absence of the pseudogap resulting from the spin density wave phase, which was found to originate from the crystalline anisotropy. Magnetic moments were not spatially uniform and were sensitive to the crystal configuration. We found that a non-uniform magnetic behavior is associated with the As-Fe-As bond angle in the structure. Our findings would clarify the importance of crystal details and orbital degrees of freedom in iron-based superconductors.  
\end{abstract}

\maketitle
\section{Introduction}
 
The discovery of the superconductivity phase in LaFeAsO$_{1-x}$F$_x$ with T$_c=26$ K by Hosono et al. [1] focused attention towards the possible relationships between the superconductivity and the magnetism. The role of magnetism and its underlying mechanism are yet to be understood. Recently, the emergence of a second antiferromagnetic (AF2) phase in the heavily H-doped LaFeAsO$_{1-x}$H$_x$  $(x\sim0.5 )$ has been reported [2,3], after the emergence of the double-dome-structure superconducting phase in these materials [4]. The origin of the phenomena, i.e. AF2 and the double-dome structure, is still under debate. Some theoretical studies have attempted to discover the origin of the phenomena [3-9]. Some studies have considered the Fermi surface (FS) nesting properties in the itinerant picture [3-5]. The orbital and spin fluctuations have been both candidates for the origin of the AF and SC phases in the materials. The spin fluctuation enhancement due to the next-nearest-neighbor (diagonal) hoppings, which originate from the disconnected FS, has been debated to create the double-dome structure [5]. The non-nematic orbital fluctuation, which induces the anion height instability, has been mentioned as a key to the understanding of the phenomena [6]. However, as the non-negligible role of the electronic correlation of high-Tc iron-based superconductors has been pointed out [10-16], considering localized aspects seems necessary. By using the DFT+DMFT method, Moon et al.  [8] found that the d$_{xy}$ orbital becomes the dominant electron hopping channel with an increased electronic correlation and an increased magnetic strength, due to structural modifications, for high doping levels in the second AF phase. The finding suggests that the indispensable role of the electronic correlation and the detailed atomic structure must not be ignored in these materials.

Recently, the experimental evidence of  T$_c$ enhancement without the influence of spin fluctuations on LaFeAsO$_{1-x}$H$_x$ under pressure has been reported by Kawaguchi et al. [17]. Furthermore, multiple-pseudogap phases with different origins of pseudogaps have been observed in the low- and high-$x$ regions of H-doped LaFeAsO$_{1-x}$H$_x$ [18]. The findings emphasize that these two AF phases, which sandwich the double-dome-structure superconducting phase, have different origins. Therefore, besides the spin fluctuation, which has been widely accepted as the most probable candidate for the pairing glue for superconductivity [19,20], the orbital degrees of freedom have been argued as the origin of the pairing mechanism in these materials [9,21-23]. Meanwhile, experimental [24-26] and theoretical [23,27] evidence of the role of the orbital degrees of freedom in these materials has been growing. For example, it has been discussed that the nematicity order in FeSe at a low temperature is due to the orbital degrees of freedom [28,29]. However, the evidence of the in-plane electronic anisotropy in the heavily H-doped LaFeAsO$_{1-x}$H$_x$ [3,30] increases the probability of the influence of the orbital degrees of freedom in the compound. Figuring out the origin of the in-plane electronic anisotropy could help in finding possible pairing mechanisms in the materials. 

\begin{figure*}
\centerline{\includegraphics[width=1.0\linewidth]{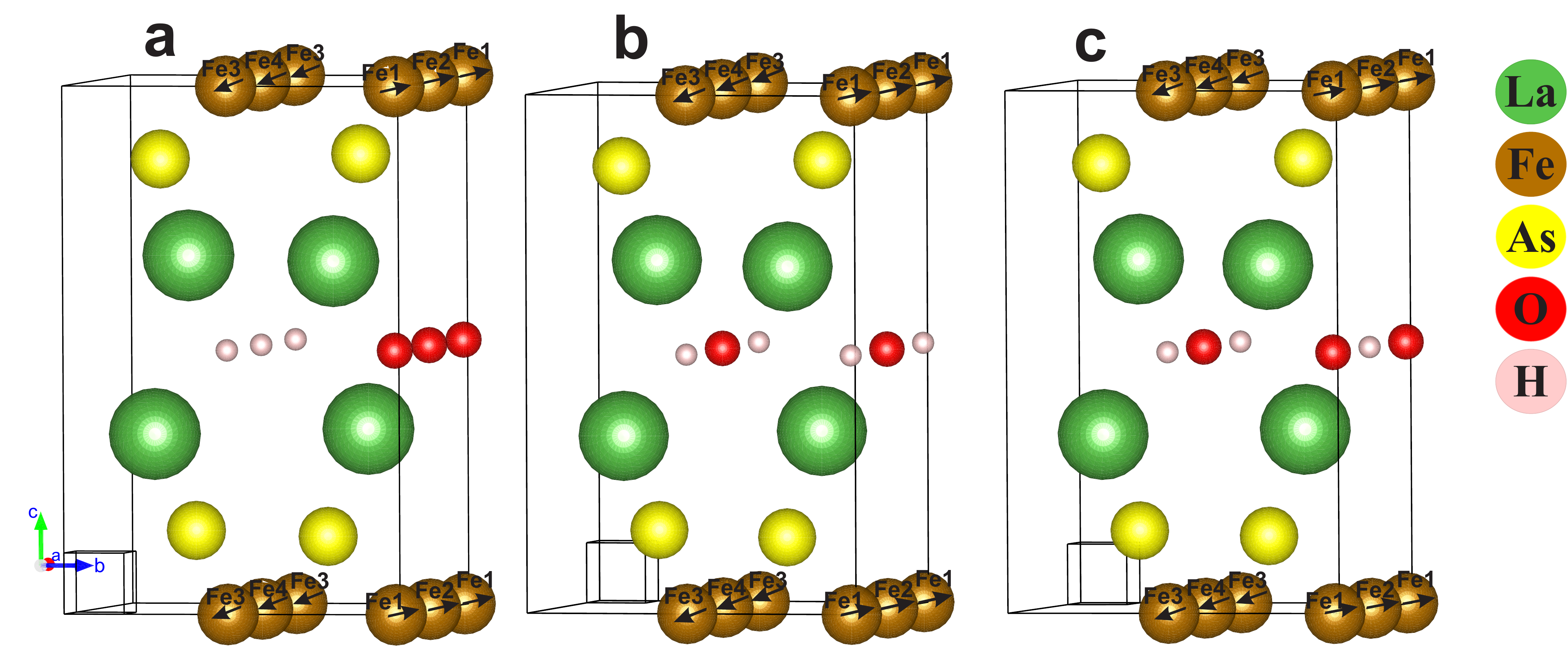}}
\caption{(Color online) Three different AF order configurations according to O and H positions: (a) AF1 (O1, in the paramagnetic state), (b) AF2 (O2, in the paramagnetic state), and (c) AF3 (O3, in the paramagnetic state). }
\end{figure*}

The strength of the electronic correlation and its role in the magnetism and the superconductivity of iron-based superconductors are still under debate. A wide range of correlations, from weak [19,31,32] to strong [10,33], have been applied to these materials in theoretical studies. A moderate correlation region, where electrons are neither fully itinerant nor fully localized, best describes their multi-orbital nature. An intriguing theory proposed by Haule and Kotliar [34] states that the Hund's rule coupling is indeed responsible for the correlation, which has a good agreement with the reported experimental results [12,16]. The density functional theory combined with the dynamical mean field theory (DFT+DMFT) method is very successful in predicting magnetic, structural, superconductivity, and electronic properties of these systems [11,12,16,35]. The DFT+DMFT method emphasizes the crucial role of the correlation in these materials.

Here, the second AF phase in the heavily H-doped  LaFeAsO$_{1-x}$H$_x$  $(x\sim0.5 )$ with the stripe-like order of iron moments, as reported in  [2], is explored. Three different AF configurations were determined for the orthorhombic phase relevant to the geometry of the system. These AF phases and the paramagnetic tetragonal phase in the heavily H-doped compound are investigated in the framework of the density functional theory combined with the dynamical mean field theory (DFT+DMFT) and are discussed in detail. We found a stripe-like-ordered structure of H and O atoms in  LaFeAsO$_{1-x}$H$_x$  $(x\sim0.5 )$ as the ground state in the magnetic state. This shows the strong correlation between the magnetic properties and the geometries of the systems. A Lifshitz transition was observed across the paramagnetic tetragonal phase, indicating that there is no hole pocket at the Fermi surface. the calculated magnetic moments reveal a spatially non-uniform magnetic behavior in the system at the ground state, which is confirmed by the reported experimental results [2,3,30].

\section{COMPUTATIONAL DETAILS}

The DFT+DMFT calculations were done for the paramagnetic tetragonal (T) phase and the paramagnetic orthorhombic AF (Aem2) phase, where the AF structure has a stripe-type  order of iron moments, as determined in  [2]. For the T phase, we substituted one of the O sites with H and then relaxed the structure. For the orthorhombic phase, we used a $\sqrt2$ a$\times\sqrt2$b$\times$c supercell containing four chemical formulas. The unit cell contains four oxygen atoms, two of which are replaced with hydrogen atoms. As shown in ‎Figure 1, three types of the AF order are formed. In AF1 (O1, as the paramagnetic state), the positions of H and O atoms construct a stripe-type configuration, exactly like the Fe magnetic moment order (‎Figure 1 (a)). If we switch the positions of H and O atoms in ‎Figure 1 (a), a new configuration would be generated. Therefore, AF1 has two different configurations according to the direction of the Fe moments, which are located in two rows above and below the H atoms. However, the two configurations are energetically the same. These configurations affect the magnetic moments of Fe atoms. The magnetic moments of Fe atoms (Fe3,4), which are located in two rows above and below H atoms (see ‎Figure 1 (a) (left)), are smaller in magnitude than those located in two rows above and below O atoms (Fe1,2) (see ‎Figure 1 (a) (right)). We will discuss this in the following. In AF2 (O2, as the paramagnetic state), a stripe-type configuration of H and O atoms is again formed, but the new configuration is perpendicular to the Fe moments order (‎Figure 1 (b)). In AF3 (O3, as the paramagnetic state) (‎Figure 1 (c)), the H and O atoms are arranged with a checkerboard-type order. We investigate the three configuration types at the paramagnetic and AF states.

We used the charge self-consistent combination of the density functional theory and the dynamical mean field theory (DFT+DMFT) [36], as implemented in two full-potential methods: the augmented plane-wave method and the linear muffin-tin orbital method (as in the WIEN2k package  [37]). The DFT calculations were performed in the WIEN2k package in its generalized gradient approximation [Perdew-Burke-Ernzerhof (PBE)-GGA]  [38]. The self-energy was added to the DFT results, and DMFT was applied to them. The computations were converged with respect to the charge density, total energy, and self-energy. The continuous-time quantum Monte Carlo (CTQMC) method  [39,40] in its fully rotationally-invariant form was used to solve the quantum impurity problem. All 3d orbitals, i.e. d$_{3z2-r2}$ (z2), d$_{x2-y2}$ (x2y2), d$_{xz}$ (xz), d$_{yz}$ (yz), and d$_{xy}$ (xy), were considered as correlated for the Fe site. The self-energy was analytically extended to the real axis using the maximum entropy method. 

Experimental lattice structures of the heavily H-doped LaFeAsO$_{1-x}$H$_x$  $(x\sim0.5 )$  were taken from Hiraishi et al. [2]. The structural positions were relaxed by DFT+DMFT, as reported by Haule and Pascut [35], until the force on each atom becomes less than 0.01 eV/${\AA}$, at the paramagnetic state. The same Coulomb interaction of $U =5.0$ eV and the same Hund's coupling of  $J_H =0.80$ eV were used in all of our DFT+DMFT calculations. Three fine meshes consisting of 1000, 500, and 250 k-points were employed for T, O1, and O2/O3 configurations, respectively, and a total of 3$\times$10$^9$ Monte Carlo steps were considered for each iteration for all states at 120 K.

\begin{table}
\caption{Total energies and the comparisons between them for the heavily H-doped LaFeAsO$_{1-x}$H$_x$  $(x\sim0.5 )$ at different states.}
\begin{tabular}{l*{3}{c}}
\hline
State             & Total Energy (Ry/atom)&Energy (meV/atom) \\
\hline
T &-6036.2996&42  \\
O1&-6036.3016 & 15.5  \\
O2&-6036.2998 &40\\
O3 &-6036.3020 & 9.5  \\
AF1 &-6036.3027 & 0  \\
AF2&-6036.3017& 13.5  \\
AF3&-6016.3017& 14  \\
\hline
\end{tabular}

\end{table}

\section{RESULTS AND DISCUSSION }

The results obtained for total energies and the comparisons between them for different states are reported in ‎Table 1. The ground state was found as AF1, which is energetically more favorable than AF2 and AF3 by 13.5 and 14 meV/Fe, respectively. In contrast to the AF states, O3 is the more stable state among paramagnetic states. It is interesting that for O3, the paramagnetic state has a higher energy than the AF state, meaning that the O3 configuration cannot be at its AF state. This suggests that a stripe-like configuration of O and H atoms is essential for the emergence of an AF state in the heavily H-doped  LaFeAsO$_{1-x}$H$_x$  $(x\sim0.5 )$. However, it is very interesting that AF states favor the stripe-like configuration of O and H atoms to their checkerboard-like configuration. Our calculations demonstrate that the crystalline anisotropy plays an important role in the emergence of a second AF in the heavily H-doped  LaFeAsO$_{1-x}$H$_x$  $(x\sim0.5 )$. In the following, we will focus on the physical characteristics of the AF1 and T states.

In these materials, the structural parameters, i.e. the anion height (the distance between As and the iron plane) and the As-Fe-As bond angle ($\angle (As-Fe-As)$ ), affect severely the physical characteristics. Therefore, calculating the parameters with enough accuracy is very important to reveal the physical characteristics. The paramagnetic relaxations by DFT+DMFT yield 1.43 and 1.42 ${\AA}$ for the anion height in O1 and T phases, respectively, while an experimental measurement [2] determined 1.40 and 1.41 ${\AA}$ for orthorhombic and tetragonal phases, respectively. The DFT+DMFT results show a very good accuracy, in contrast to the non-magnetic (NM) DFT that misestimates the anion heights in O1 and T phases as 1.23 and 1.24 ${\AA}$, respectively. The anion height could severely affect magnetic and superconductivity behaviors  [41,42]. Another important parameter in these materials is the interlayer distance (the distance between the FeAs and LaO layers). In our relaxations by DFT+DMFT, the interlayer distance was obtained 1.51 and 1.54 ${\AA}$ for O1 and T phases, respectively, showing a very good agreement with the experimentally obtained results of 1.53 and 1.54 ${\AA}$ for orthorhombic and tetragonal phases, respectively  [2]. The relaxations by NM-DFT overestimate the interlayer distance for O1 and T phases as 1.67 and 1.68 ${\AA}$, respectively. Our calculations would clarify that DFT+DMFT is a very skillful method in predicting the structural parameters in iron-based superconductors. 

The interlayer distance decreases from 1.81 to 1.54 ${\AA}$ when going from the undoped LaFeAsO [1] to the heavily H-doped  LaFeAsO$_{1-x}$H$_x$  $(x\sim0.5 )$ [2]. The 15 $\%$  decrease makes the charge transfer between the layers much easier, severely affecting the superconductivity transition temperature (T$_c$) and increasing T$_c$ to its optimum value. The increased charge transfer in the heavily H-doped  LaFeAsO$_{1-x}$H$_x$ could explain the higher T$_c$ values in the second dome of the double-dome structure in these compounds. On the other hand, the anion height increases from 1.32 to 1.41 ${\AA}$ when going from the undoped LaFeAsO [1] to the heavily H-doped  LaFeAsO$_{1-x}$H$_x$  $(x\sim0.5 )$ [2].Therfore, the intralayer parameter, i.e. the anion height, can affect the magnetic and superconductivity behaviors in iron-based compounds [41,42]. The effects could be induced by the electronic correlation [8], which is highly affected by the anion height [43]. However, the proximity of the second dome with its higher  T$_c$ to a second AF state makes it clear that the structural parameters play an important role in the physical characterization of the heavily H-doped LaFeAsO$_{1-x}$H$_x$  $(x\sim0.5 )$. 

Magnetic calculations show a non-uniform magnetic behavior in AF1, which is induced by the crystalline anisotropy. The geometry of the system on the blocking-layer (LaO/H), which is a stripe-like-ordered structure of O and H atoms, affects the magnitude of the Fe magnetic moment. The magnetic moments of Fe atoms (Fe3,4) located in two rows above and below H atoms (see ‎Figure 1 (a) (left)), are smaller in magnitude than those of Fe atoms (Fe1,2) located in two rows above and below O atoms, (see ‎Figure 1 (a) (right)). We obtained 1.6 and 1.1 $\mu_B$ for the magnetic moments of Fe1,2 and Fe3,4, respectively. The difference between the magnetic moments is 0.5  $\mu_B$, which is significant. To clarify the phenomenon, we calculated the bond angle of those types of Fe atoms that have different magnetic behaviors.

\begin{equation}
\begin{array}{llll}
\Delta \angle (As-Fe-As)= \\
 \angle (As-Fe3,4-As)-\angle (As-Fe1,2-As)=1.202{}^\text{o} \\
 \angle (As-Fe3,4-As)=108.762{}^\text{o};\\
\angle (As-Fe1,2-As)=107.560{}^\text{o}
\end{array}
\end{equation}

\begin{figure}
\centerline{\includegraphics[width=0.85\linewidth]{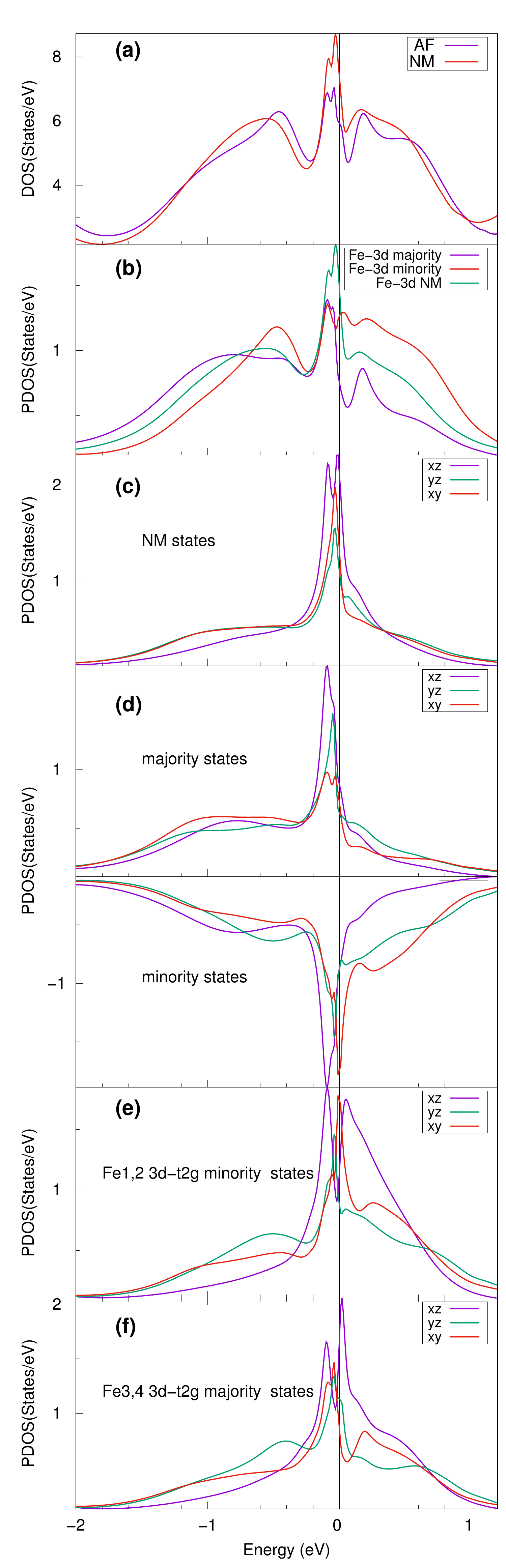}}
\caption{(Color online) Density of states (DOS) of the O1 configuration (see ‎Figure 1 (a)) at the paramagnetic (non-magnetic (NM)) and AF1 states in the heavily H-doped LaFeAsO$_{1-x}$H$_x$  $(x\sim0.5 )$: (a) the total DOS and (b) the Fe-3d projected DOS. The xy, xz, and yz PDOSs in (c) NM, (d) AF1, (e) Fe1,2 (AF1), and (f) Fe3,4 (AF1). }
\end{figure}

We found the difference of 1.2${}^\text{o}$ between the bond angles of those types of Fe atoms that have different magnetic behaviors. The different bond angles cause a huge difference between the magnetic moments of these different types of Fe atoms. The phenomenon shows directly the effects of the bond angle on the magnetic behaviors of the materials. This non-uniform magnetic behavior is consistent with the existing experimental measurements [2,3,30]. This suggests that the bond angle is the origin of the non-uniform magnetic behavior in the heavily H-doped LaFeAsO$_{1-x}$H$_x$  $(x\sim0.5 )$. 

\begin{figure*}
\rightline{\includegraphics[width=0.9\linewidth]{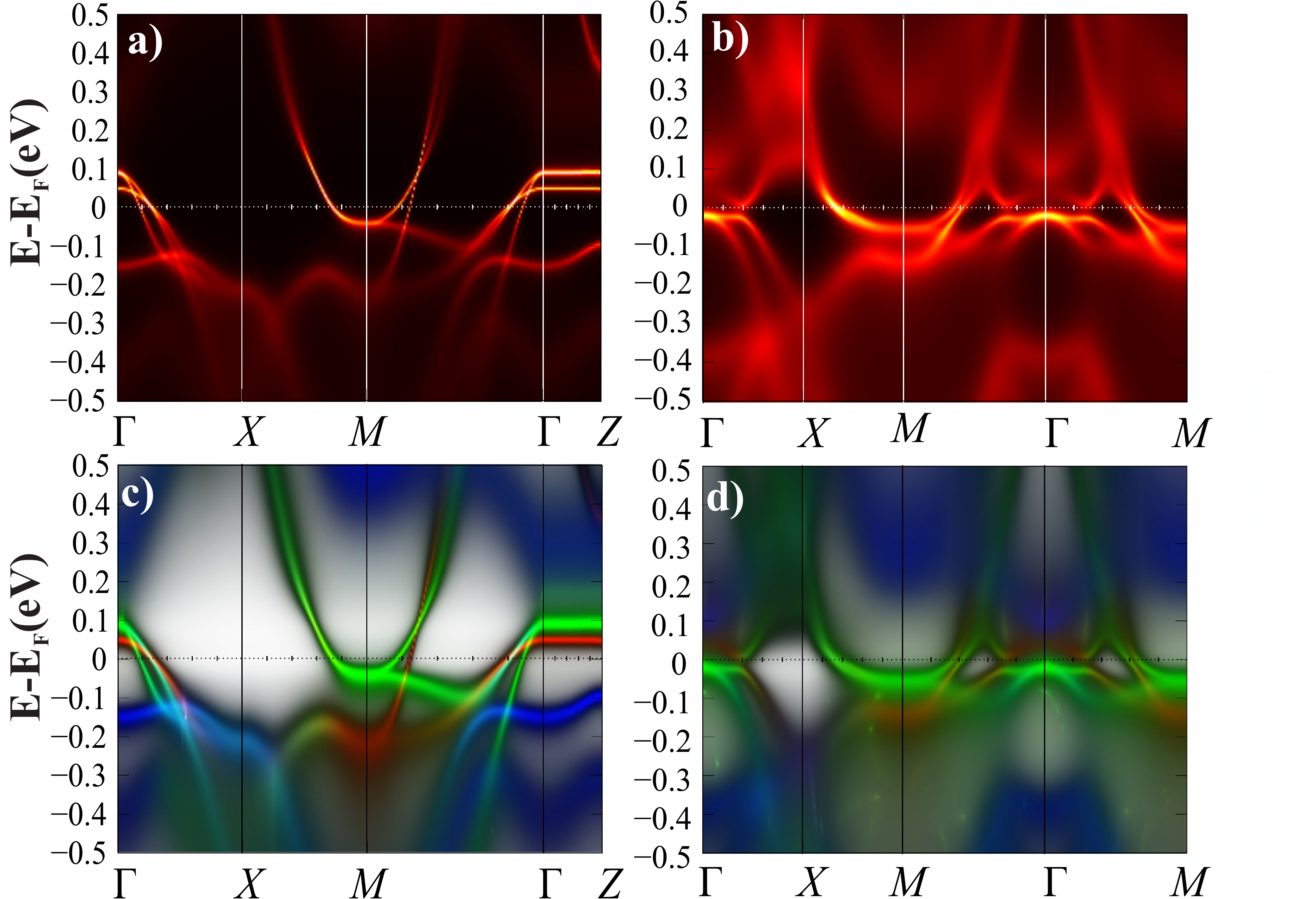}}
\caption{(Color online). The spectral functions, computed by DFT+DMFT, for the paramagnetic state of (a) the undoped LaFeAsO and (b) the heavily H-doped  LaFeAsO$_{1-x}$H$_x$  $(x\sim0.5 )$ (T phase), and their corresponding orbital resolved spectral functions for (c) the undoped LaFeAsO and (d) the heavily H-doped  LaFeAsO$_{1-x}$H$_x$  $(x\sim0.5 )$ (z2 and x2y2 are in blue, xz and yz are in green, and xy is in red).}
\end{figure*}

To demonstrate the effects of the electronic structure on physical characteristics of the heavily H-doped LaFeAsO$_{1-x}$H$_x$, its localized spectral functions are plotted in ‎Figure 2 in detail. The total density of states (DOS) shows that three peaks appear around the Fermi energy (E$_F$) (‎Figure 2 (a)), which signifies the Fermi liquid behavior. For the paramagnetic (NM) state, the middle peak shifts by $\sim50$ meV to below E$_F$, while the upper and lower peaks are at 100 and -240 meV, respectively. For AF1, the peaks shift toward  E$_F$ and no spin density wave (SDW) pseudogap can be seen at  E$_F$, and hence the AF1 state has a major difference from the AF state at the undoped LaFeAsO [8]. The absence of an SDW pseudogap has been previously confirmed theoretically [7] and experimentally [17,18]. As shown in ‎Figure 2 (b), the total DOS around  E$_F$ gains its weight mainly from Fe-3d orbitals. Therefore, we investigated the projected density of states (PDOS) of Fe-3d-t2g (xz, yz, and xy) orbitals that have higher energy levels and have the main contributions to DOS at  E$_F$ (N$_{E_F}$). Our calculations showed that even at the paramagnetic state (‎Figure 2 (c)), xz and yz orbitals split and they are no more degenerate. This is due to the crystalline anisotropy in the system that destroys the C$^4$ symmetry by forming a stripe-like configuration of H and O atoms. This crystalline anisotropy induces an orbital anisotropy is the iron plane in the heavily H-doped  LaFeAsO$_{1-x}$H$_x$  $(x\sim0.5 )$, which increases the contribution of the xz orbital at N$_{E_F}$. Our results could explain the in-plane electronic anisotropy revealed by NMR measurements [3,30] in the heavily H-doped LaFeAsO$_{1-x}$H$_x$. The majority and the minority states of AF1 for xz, yz, and xy orbitals are plotted in the top and bottom panels in ‎Figure 2 (d). The crystalline anisotropy increases the contribution of the xz orbital to the magnetic moment, while decreases the contribution of the yz orbital. Nevertheless, besides increasing the contribution of the xz orbital induced by the crystalline anisotropy, the xy orbital plays a major role in the local magnetic moment. As shown in ‎Figure 2 (d), the minority states of the xy orbital were nearly half-filled, meaning that the xy orbital becomes a major candidate for the dominant electron hopping channel in the heavily H-doped LaFeAsO$_{1-x}$H$_x$  $(x\sim0.5 )$. To understand the variation of the magnetic moment between two adjacent rows of Fe atoms, we calculated the majority states of Fe1,2 and the minority states of Fe3,4 for xz, yz, and xy orbital at the AF1 state, as shown in ‎Figure 2 (e) and 2 (f). Our results show that all the three orbitals contribute to the variation in the magnetic moment. The contribution from the xy orbital is surprising because the angle bond mostly affects the xz and yz orbitals. This highlights the role of the xy orbital in the electron hopping and suggests the major role of the xy orbital in the magnetic behaviors of LaFeAsO for high levels of H-doping. Our calculations emphasize the important role played by the structural details and the orbital degrees of freedom in the physical characteristics of iron-based superconductors.

The spectral functions, computed by DFT+DMFT, for the paramagnetic state of  LaFeAsO$_{1-x}$H$_x$   at $x =0$ and $0.5 $ (T phase) and their corresponding orbital resolved spectral functions are plotted in ‎‎Figure 3. For $x=0$, we used the structural parameter reported by Kamihara et al.  [1]. The spectral function of the undoped compound shows three hole pockets at the center of the Brillouin zone ($\Gamma$) and two electron pockets at the zone corner ($M$), consistent with the previous DFT+DMFT studies  [8,10]. At the center, the outer band has mostly the xy character. The middle band is a combination of xz/yz and x2y2 characters, and the inner band is mostly composed of the xz/yz character, as it is illustrated in ‎Figure 3 (c). At the zone corner, two elliptical electron pockets create the Fermi surface (see ‎Figure 4 (a)). The outer band has mostly the xy character and the inner band has the xz/yz character. The substitution of H atoms into the half of the O sites was performed as electron-doping. This enlarged the electron pockets at the zone corner and destroyed the hole pockets at the zone center (see ‎Figure 4 (b)). Our calculations predict a Lifshitz transition in the topology of the Fermi surface through the substitution of H atoms into the half of the O sites. This phenomenon occurs when an electron-like band with the zx/yz character approaches  E$_F$ at $\Gamma$ and crosses the hole-like band with the xy character at E$_F$ (see ‎Figure 3 (d)). The Fermi surface without hole pockets with a high Tc has been reported in iron chalcogenides, i.e. K$_y$Fe$_2$Se$_{2-x}$ [44] and the monolayer FeSe over the SrTiO$_3$ substrate [45]. However, in iron pnictides, the superconductivity is suppressed due to the heavily electron-doping, i.e. in BaFe$_{2-x}$Co$_x$As$_2$ [46]. However, the coexistence of electron- and hole-like bands with their xz/yz and xy characters, respectively, demonstrates that the xz/yz orbitals are more filled than the xy orbital. This emphasizes the role of the xy orbital as the dominant electron hopping channel for the heavily H-doped  LaFeAsO$_{1-x}$H$_x$  $(x\sim0.5 )$. Moon et al. [8] found the same behavior for the xy orbital in the framework of DFT+DMFT. These results would clarify the major role of the xy orbital in iron-based superconductors.

The schematics of the Fermi surface for the undoped LaFeAsO and the heavily H-doped  LaFeAsO$_{1-x}$H$_x$  $(x\sim0.5 )$, derived from the spectral functions in ‎Figure 3, are illustrated in ‎Figure 4. The colors show the orbitals that mostly characterize the pockets. As discussed before, the heavily H-doping behaves as a high level of electron-doping and forms two very wide electron pockets at the zone corner, while destroys the hole pockets at the zone center. The transition in the topology of the Fermi surface during the process is mainly induced by the filling of the xz/yz orbitals, which creates an electron-like band at the zone center. Nevertheless, the xy orbital, which is less filled than the xz/yz orbitals and creates the hole-like band at the zone center,   has also an important role in this phenomenon. The Hund's coupling $J_H$, which is the origin of the differences among the correlation strengths of different orbitals, plays an important role in these phenomena. This suggests the Hund's metal behavior, predicted by Haule and Kotlair [35], for these materials.

\begin{figure}
\centerline{\includegraphics[width=1.0\linewidth]{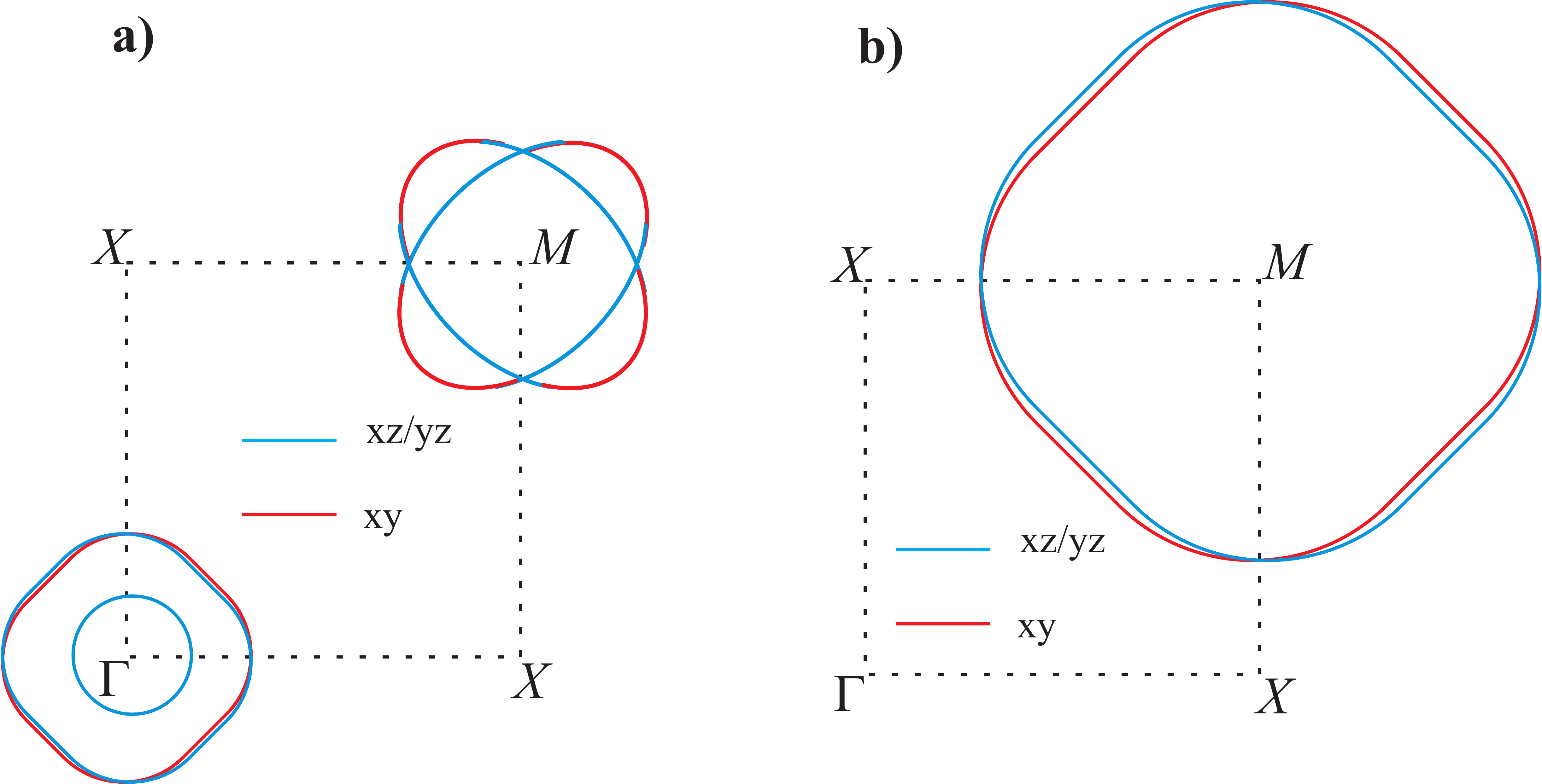}}
\caption{(Color online) Schematics of the Fermi surface for (a) the undoped LaFeAsO and (b) the heavily H-doped LaFeAsO$_{1-x}$H$_x$  $(x\sim0.5 )$.  
}

\end{figure}

In the heavily H-doped  LaFeAsO$_{1-x}$H$_x$  $(x\sim0.5 )$, there are some structural variations as compared to the undoped LaFeAsO that we must notice them. Firstly, the interlayer distance between FeAs and LaO/H layers is shorter, which enhances the charge transfer between the layers. The decrease in the interlayer distance occurs when applying pressure on LaFeAso [43,47]. This could help the system to reach its maximum T$_c$ that happened at $x\sim0.36$ at the top points of the second-dome. Furthermore, the charge transfer enhancement could destroy the magnetic state in the system [1,47,48]. Secondly, the intralayer distance, i.e. the anion height, has increased, which is in contrast to its behavior under pressure  [43,47]. This severely affects the electronic correlation [8,43] and the magnetic moments [41]. An increase in the anion height increases the correlation in Fe-3d orbitals and can strengthen the magnetic moments. However, due to the increased anion height in our calculations, the Fe-3d orbitals gain some weight and become more localized than the undoped state [8], which results in a larger magnetic moment for the second AF. The duality between the increased anion height and the decreased interlayer distance could create a competitive behavior that depends on the one that is able to overcome. It could clear the duality between the high-temperature superconductivity and AF phases in the heavily H-doped  LaFeAsO$_{1-x}$H$_x$  $(x\sim0.5 )$. 

The stripe-like-ordered structure of H and O atoms as the ground state for the same AF phase is very interesting. This configuration could explain some rare experimental results such as the in-plane electronic anisotropy and the non-uniform magnetic behavior [3,30]. Our calculations showed the essential role of the crystalline anisotropy induced by this configuration in the second AF. Song et al. [49] reported the experimental stripe-like order of Cu and Fe atoms in NaFe$_{1-x}$Cu$_x$As $(x\sim0.5 )$. It is suggested that the new configuration, found by our calculations, be studied experimentally. Our finding could propose a new route for studying the second AF phase in the heavily doped compound of LaFeAs$_{1-x}$P$_x$O $(x\sim0.5 )$ [50,51].

\section{Conclusion}
Electronic and magnetic characteristics of the heavily H-doped  LaFeAsO$_{1-x}$H$_x$  $(x\sim0.5 )$ were studied by the DFT+DMFT method. We found a stripe-like configuration of H and O atoms, similar to the AF order, as the ground state. Our calculations showed that the new configuration could explain experimental results, i.e. the non-uniform magnetic behavior and the in-plane electronic anisotropy. The phenomena originate from the crystalline anisotropy, which plays an important role in the heavily H-doped  LaFeAsO$_{1-x}$H$_x$  $(x\sim0.5 )$. The non-uniform magnetic behavior is caused by the difference between the bond angles of various types of Fe atoms with different magnetic moments. We also found a significant anisotropy between the xz and yz orbitals in this compound without any pseudogap at  E$_F$, indicating that it is originated from the crystalline anisotropy. A transition in the topology of the Fermi surface was observed in the heavily H-doped  LaFeAsO$_{1-x}$H$_x$  $(x\sim0.5 )$ in the T phase, which originates from the different correlation strengths of different orbitals. The observation would clarify the major role of the Hund's coupling $J_H$ in this compound. Our results highlight the role of the structural details and the electronic correlation in iron pnictides.

\section*{Acknowledgement}
We are very grateful to the Professor Kristjan Haule for providing the DFT+DMFT code and also his tutorials and responses to our questions and comments. We also would thank Dr. Gheorghe L. Pascut for his tutorials and comments. The calculation was performed in the high-performance computing (HPC) Center of Tarbiat Modares University (TMU).

\end{document}